\documentclass[onecolumn,showpacs,preprintnumbers,amsmath,amssymb,prb,floatfix]{revtex4}

\usepackage{graphicx}
\usepackage{dcolumn}
\usepackage{bm}
\usepackage{color}

\begin{document}

\preprint{APS/123-QED}

\title{Electron transport through Aharonov-Bohm interferometer\\with laterally coupled double quantum dots}

\author{Toshihiro Kubo$^{1,}$}
 \email{kubo@tarucha.jst.go.jp}
\author{Yasuhiro Tokura$^{1,2}$}
\author{Tsuyoshi Hatano$^{1}$}
\author{Seigo Tarucha$^{1,3}$}%
 \affiliation{%
$^1$Quantum Spin Information Project, ICORP, JST, Atsugi-shi, Kanagawa 243-0198, Japan\\
$^2$NTT Basic Research Laboratories, NTT Corporation, Atsugi-shi, Kanagawa 243-0198, Japan\\
$^3$Department of Applied Physics, University of Tokyo, Hongo, Bunkyo-ku, Tokyo 113-0033, Japan
}%

\date{\today}

\begin{abstract}
We theoretically investigate electron transport through an Aharonov-Bohm interferometer containing laterally coupled double quantum dots. We introduce the indirect coupling parameter $\alpha$, which characterizes the strength of the coupling via the reservoirs between two quantum dots. $|\alpha|=1$ indicates the strongest coupling, where only a single mode contributes to the transport in the system. Two conduction modes exist in a system where $|\alpha|\neq 1$. The interference effects such as the Fano resonance and the Aharonov-Bohm oscillation are suppressed as the absolute value of the parameter $\alpha$ decreases from $1$. The linear conductance does not depend on the flux when $\alpha=0$ since it corresponds to independent coupling of the dots to the reservoir modes.
\end{abstract}

\maketitle

\section{\label{introduction}Introduction}
Quantum phase coherence in mesoscopic systems has attracted the attention of many physicists. Quantum phase coherence is detectable by interference experiments. In interference experiments with an Aharonov-Bohm (AB) ring containing a quantum dot (QD), quasi-periodic modulation of the tunneling current as a function of the magnetic flux through the ring has been experimentally demonstrated\cite{yacoby,schuster,ji}. This reflects the fact that the phase coherence is maintained during the tunneling process through a QD. The Fano effect\cite{fe} is another important interference effect in mesoscopic physics. The Fano effect occurs in a system in which discrete and continuum energy states coexist\cite{kobayashi,fe}.

Recently the AB oscillations of a tunneling current passing through a laterally coupled double quantum dot (DQD) system were observed by Holleitner et al. using a lateral DQD\cite{holleitner}. Such oscillations were also observed by a group including two of the present authors using a vertical DQD\cite{hatano} in a weak inter-dot tunnel coupling regime. Motivated by these experimental results, electron transport through such a system has been investigated theoretically\cite{kubala,kang,bai}. Moreover, such systems are very interesting since it has been theoretically predicted that cotunneling currents passing through spin-singlet and triplet states have different AB oscillation phases\cite{loss}. DQD has been attracting attention as an important device structure for entangled spin qubit operations\cite{loss2,hatano2,petta}. 

In this paper, we consider the transport through an AB interferometer containing a laterally coupled DQD. Although the electron spin and interaction effects are crucial in the previous theoretical proposals, here we disregard them and focus on the single particle interference properties. Instead, we introduce the indirect coupling parameter $\alpha$, which characterizes the strength of the coupling via the reservoirs between two QDs\cite{raikh}. A system with a maximum coupling $|\alpha|=1$ has already been widely studied theoretically\cite{kubala,kang,bai}. In actual systems, however, such a case is very special and most experimental situations correspond to $|\alpha|<1$. We found that the number of conduction modes is two except when $|\alpha|=1$, where only a single mode contributes to the transport. The situation where $\alpha=0$ has also been explored in the context of the orbital Kondo problem\cite{wilhelm,orbital}. We calculate the tunneling current through the DQD systems in terms of Green's function techniques for non-interacting systems\cite{MW,JWM}. By using an exact expression of the tunneling current, we examine AB oscillations through DQD from the weak inter-dot tunnel coupling regime to the strong inter-dot tunnel coupling regime. The visibility of AB oscillation (i.e., the ratio of the AB oscillation amplitude and the current maximum) decreases as the absolute value of the indirect coupling parameter $\alpha$ decreases. The current is independent of the flux at $\alpha=0$, and the visibility becomes zero. As a function of energy, the linear conductance indicates an asymmetric Fano lineshape when $|\alpha|=1$,  when the energy of the (anti-)symmetric state is near the Fermi level of the reservoirs. However, the Fano resonance is suppressed as $|\alpha|$ decreases and finally the linear conductance exhibits two peaks at $\alpha=0$ corresponding to the Breit-Wigner resonances by the symmetric and anti-symmetric states.

This paper is organized as follows. In Sec. \ref{model}, a standard tunneling Hamiltonian is employed to describe an AB interferometer containing a laterally coupled DQD. We introduce the indirect coupling parameter $\alpha$ in Sec. \ref{icp}. We derive the linear conductance at zero temperature in Sec. \ref{lc}. In Sec. \ref{linear}, we provide the energy dependences of the linear conductance at zero temperature and we show that the Fano resonance is suppressed in a system where $|\alpha|\neq1$. In Sec. \ref{ab}, we discuss the AB oscillations of the linear conductance at zero temperature. In Sec. \ref{asymmetry}, we extend the argument where the magnetic flux and tunnel coupling between reservoirs and QDs  are asymmetric. All our results are summarized in Sec. \ref{conclusion}. In Appendix \ref{dicp}, we give the detailed derivation of the indirect coupling parameter $\alpha$ for the two-dimensional and three-dimensional reservoirs.  In Appendix \ref{transmission}, by diagonalizing the transmission matrix, we show that there are two conduction modes when $|\alpha|\neq1$. However, there is only a single conduction mode when $|\alpha|=1$. When $\alpha=0$, the two conduction modes correspond to the Breit-Wigner resonances by the symmetric and anti-symmetric states.

\section{\label{model}Model}
\begin{figure}
\includegraphics[scale=0.5]{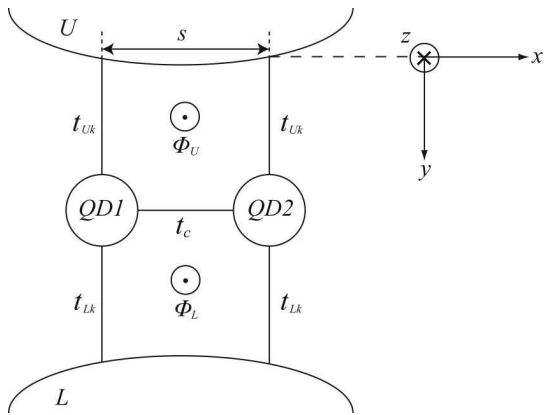}
\caption{\label{fig:system} Schematic diagram of an AB interferometer containing a laterally coupled DQD. The magnetic fluxes threading the upper and lower subcircuits are $\Phi_U$ and $\Phi_L$, respectively, and cause the AB effect. $s$ is the distance between the two QDs.}
\end{figure}

We consider an AB interferometer containing a DQD coupled to two reservoirs as shown in Fig. \ref{fig:system}. Only a single energy level in each QD is assumed to be relevant and we ignore the spin degree of freedom.  We model this system with the following Hamiltonian:
\begin{equation}
H=H_R+H_{DQD}+H_T,\label{ham1}
\end{equation}
where $H_R$ describes Fermi seas of non-interacting electrons in two reservoirs
\begin{equation}
H_R=\sum_{\nu\in\{U,L \}}\sum_k\epsilon_{\nu k}{c_{\nu k}}^{\dagger}c_{\nu k}.
\end{equation}
Here $\epsilon_{\nu k}$ is the energy of conduction electrons with wave number $k$ in the reservoirs $\nu=U,L$ and the operator $c_{\nu k}$ $({c_{\nu k}}^{\dagger})$ annihilates (creates) an electron in reservoirs. $H_{DQD}$ is the Hamiltonian of the isolated DQD
\begin{eqnarray}
H_{DQD}=\sum_{i=1}^2\epsilon_0{d_i}^{\dagger}d_i+t_c\left({d_1}^{\dagger}d_2+h.c. \right).
\end{eqnarray}
Here $\epsilon_0$ is the single-particle energy level of two QDs\cite{offset} and $d_i$ $({d_i}^{\dagger})$ annihilates (creates) an electron in the $i$th QD ($i=1,2$). The second term represents direct tunneling between two dots. The energy levels of the dots split to form symmetric and anti-symmetric states because of inter-dot tunnel coupling, $t_c$. Their energy levels are $\epsilon_s=\epsilon_0+t_c$ and $\epsilon_a=\epsilon_0-t_c$, where we choose the gauge such that $t_c$ is real and negative. $H_T$ describes the tunneling Hamiltonian between reservoirs and QDs
\begin{eqnarray}
H_T&=&\sum_k\left[t_{Uk}^{(1)}e^{i\frac{\phi_U}{2}}{c_{U k}}^{\dagger}d_1+t_{Uk}^{(2)}e^{-i\frac{\phi_U}{2}}{c_{Uk}}^{\dagger}d_2\right.\nonumber\\
&&\left.+t_{Lk}^{(2)}e^{i\frac{\phi_L}{2}}{c_{Lk}}^{\dagger}d_2+t_{Lk}^{(1)}e^{-i\frac{\phi_L}{2}}{c_{Lk}}^{\dagger}d_1+h.c. \right]\nonumber\\
&=&\sum_{\nu\in\{U,L \}}\sum_k\sum_{i=1}^2\left[t_{\nu k}^{(i)}(\phi_{\nu}){c_{\nu k}}^{\dagger}d_i+h.c. \right],
\end{eqnarray}
where $t_{\nu k}^{(i)}$ is real. The factors $e^{\pm i\frac{\phi_{\nu}}{2}}$ indicate the effect of the magnetic flux ($\phi_{\nu}=2\pi\Phi_{\nu}/\Phi_0$ is an AB phase in each subcircuit, where $\Phi_{\nu}$ is the magnetic flux threading through each subcircuit as shown in Fig. 1, and $\Phi_0=h/e$ is the magnetic flux quantum).

\section{\label{current}Tunneling current}
For non-interacting systems, the tunneling current can be written as\cite{MW}
\begin{eqnarray}
I&=&\frac{e}{h}\int d\epsilon\left[f_U(\epsilon)-f_L(\epsilon) \right]\mbox{Tr}\left\{\bm{G}^r(\epsilon)\bm{\Gamma}^U(\epsilon)\bm{G}^a(\epsilon)\bm{\Gamma}^L(\epsilon) \right\}\nonumber\\
&=&\frac{e}{h}\int d\epsilon\left[f_U(\epsilon)-f_L(\epsilon) \right]T(\epsilon).
\end{eqnarray}
Here $f_{\nu}(\epsilon)=1/\left[1+e^{(\epsilon-\mu_{\nu})/k_BT} \right]$ with $\nu=U/L$ is the Fermi-Dirac distribution function, $\mu_{\nu}$ being the chemical potential of the reservoir $\nu$. Boldface notations indicate $2\times 2$ matrices, where $\bm{G}^{r(a)}(\epsilon)$ is the retarded (advanced) Green's function of a DQD, $\bm{\Gamma}^{U(L)}$ is the linewidth function defined by
\begin{equation}
\Gamma_{ij}^{\nu}(\epsilon)\equiv 2\pi\sum_k{t_{\nu k}^{(i)}}^*(\phi_{\nu})t_{\nu k}^{(j)}(\phi_{\nu})\delta(\epsilon-\epsilon_{\nu k}),\label{g}
\end{equation}
and $T(\epsilon)$ is the transmission probability at energy $\epsilon$.

\subsection{\label{icp}Indirect coupling parameter $\alpha$}
Here we estimate the linewidth function based on Ref. \onlinecite{stm} and focus on the three-dimensional (3D) reservoirs. The detailed derivation of the indirect coupling parameter $\alpha$ for not only 3D but also two-dimensional (2D) reservoirs is given in Appendix \ref{dicp}. According to the Bardeen's formula\cite{bardeen}, the flux independent tunneling matrix element $t_{\nu k}^{(i)}$ is given by
\begin{eqnarray}
t_{\nu k}^{(i)}=\frac{\hbar^2}{2\mu}\int_SdS\left\{\phi_{\nu k}(\vec{r})\nabla{\phi_d^{(i)}}^*(\vec{r})-{\phi_d^{(i)}}^*(\vec{r})\nabla\phi_{\nu k}(\vec{r}) \right\},\label{bardeen-form}
\end{eqnarray}
where $\mu$ is the effective mass, the 2D area $S$ for the integration is in the tunneling potential, $\phi_{\nu k}(r)$ is the wave function of evanescent mode of the reservoir $\nu$, and $\phi_d^{(i)}(r)$ is the wave function of a localized electron in the dot $i$.

\begin{figure}
\includegraphics[scale=0.5]{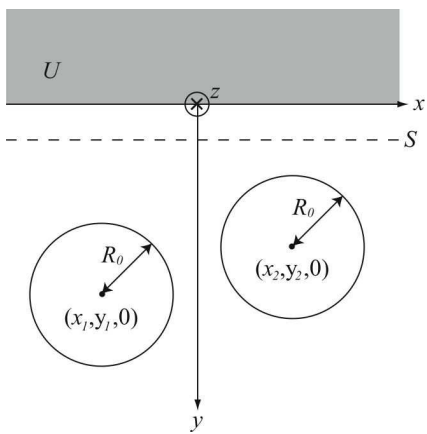}
\caption{\label{fig:system2} Model to estimate the tunneling coupling $t_{\nu k}^{(i)}$ between the reservoirs and the two dots. $S$ indicates the area in the tunneling potential. The radius of the QD is $R_0$ and the center position of $i$th dot is $(x_i,y_i,0)$.}
\end{figure}
We calculate the tunneling matrix elements for the model shown in Fig. \ref{fig:system2}, where we consider only the upper reservoir. As shown in Fig. \ref{fig:system2}, the boundary of the reservoir is a flat surface, the dot size is $R_0$, and the potential profile is
\begin{eqnarray}
U(\vec{r})=\left\{
  \begin{array}{cc}
    0   &  (\mbox{in the reservoir},\ y<0)  \\
    \sum_{i=1}^2U_{dot}(\vec{r}-\vec{r}_i)   &  (|\vec{r}-\vec{r}_i|<R_0)  \\
    U   &  (\mbox{others including the area}\ S)  \\
  \end{array}
\right.,
\end{eqnarray}
where $\vec{r}_i=(x_i,y_i,0)$ is the center of the $i$th dot. Then the wave functions of reservoir modes are
\begin{equation}
\phi_{Uk}(\vec{r})=\frac{\sqrt{2}}{L^{3/2}}e^{ik_xx+ik_zz}\sin\left(\frac{\theta}{2}\right)e^{-\kappa y},
\end{equation}
where the decay constant $\kappa$ is
\begin{eqnarray}
\kappa=\frac{1}{\hbar}\sqrt{2\mu\left(U-\frac{\hbar^2{k_y}^2}{2\mu}\right)},
\end{eqnarray}
and the scattering phase shift $\theta$ satisfies $\sin\theta=\frac{2\kappa k_y}{\kappa^2+{k_y}^2}$. $L$ is the system size and ${k_{\parallel}}^2={k_x}^2+{k_z}^2$. For the dot, we consider only the symmetric mode, namely zero angular momentum state,
\begin{equation}
\phi_d^{(i)}(r>R_0)=D_3\frac{1}{\sigma |\vec{r}-\vec{r}_i|}e^{-\sigma |\vec{r}-\vec{r}_i|}
\end{equation}
where $\sigma^2+{k_{\parallel}}^2=\kappa^2$ and $D_3$ is a constant related to the detail of the potential inside the dot $U_{dot}$. Using Eq. (\ref{bardeen-form}), the tunneling matrix elements are
\begin{eqnarray}
t_{Uk}^{(i)}=\frac{\hbar^2}{2\mu}\frac{\sqrt{2}}{L^{3/2}}\sin\left(\frac{\theta}{2} \right)\frac{4\pi D_3}{\sigma}e^{-\kappa y_i+ik_xx_i}.
\end{eqnarray}

Then we calculate the linewidth function $\Gamma_{ij}^U(\epsilon)$ using Eq. (\ref{g}). The diagonal elements $\Gamma_{ii}^U(\epsilon)$ are real, which are equal since we assumed $y_1=y_2=y_d$. We define the indirect coupling parameter as follows:
\begin{eqnarray}
\alpha^U(\epsilon)\equiv\frac{\Gamma_{12}^U(\epsilon)}{\sqrt{\Gamma_{11}^U(\epsilon)\Gamma_{22}^U(\epsilon)}}.
\end{eqnarray}
We found the indirect coupling parameter at the Fermi energy $\mu_U=\frac{\hbar^2{k_F}^2}{2\mu}$ with $U\gg \mu_U$,
\begin{eqnarray}
\alpha^U(\mu_U)\sim\frac{3}{k_Fs}j_1(k_Fs)\label{al1},
\end{eqnarray}
where $j_1$ is the first order spherical Bessel function and for $U\gtrsim \mu_U$
\begin{eqnarray}
\alpha^U(\mu_U)\sim\frac{(2y_d)^3}{(s^2+(2y_d)^2)^{3/2}},
\end{eqnarray}
where $s=|x_1-x_2|$. We find $|\alpha^U|\le 1$ and decreases with $s$. The oscillation with $k_Fs$ shown in Eq. (\ref{al1}) originates from the assumptions of ballistic transport in the reservoirs and the flat reservoir surface. This oscillation becomes weaker and $\alpha$ decreases monotonically with $s$ if we consider curved or rough boundaries. We expect that the above discussions can be extended into the general model as shown in Fig. \ref{fig:system}. We restrict our discussion to a symmetrically coupled DQD system, $\Gamma_{11}^{\nu}(\epsilon)=\Gamma_{22}^{\nu}(\epsilon)=\gamma_{\nu}(\epsilon)$ for simplicity. Then, using a basis of localized states, we use the linewidth functions
\begin{eqnarray}
\bm{\Gamma}^U(\epsilon)=\left(
  \begin{array}{cc}
    \gamma_U(\epsilon)   &  \alpha^U(\epsilon)\gamma_U(\epsilon)e^{-i\phi_U}  \\
    \alpha^U(\epsilon)\gamma_U(\epsilon)e^{i\phi_U}   &  \gamma_U(\epsilon)  \\
  \end{array}
\right),\ \bm{\Gamma}^L(\epsilon)=\left(
  \begin{array}{cc}
    \gamma_L(\epsilon)   &  \alpha^L(\epsilon)\gamma_L(\epsilon)e^{i\phi_L}  \\
    \alpha^L(\epsilon)\gamma_L(\epsilon)e^{-i\phi_L}   &  \gamma_L(\epsilon)  \\
  \end{array}
\right),
\end{eqnarray}
in the following argument.

\subsection{\label{lc}Linear Conductance}
We obtain the following retarded Green's function using the equation of motion,
\begin{eqnarray}
\bm{G}^r(\epsilon)&=&\left[\left( \bm{g}^r(\epsilon) \right)^{-1}+\frac{i}{2}\left(\bm{\Gamma}^U(\epsilon)+\bm{\Gamma}^L(\epsilon)\right) \right]^{-1}\nonumber\\
&=&\left(
  \begin{array}{cc}
     (\epsilon-\epsilon_0)/\hbar+i(\gamma_U(\epsilon)+\gamma_L(\epsilon))/2  &  -t_c/\hbar+i(\alpha^U(\epsilon)\gamma_U(\epsilon)e^{-i\phi_U}+\alpha^L(\epsilon)\gamma_L(\epsilon)e^{i\phi_L})/2  \\
     -t_c/\hbar+i(\alpha^U(\epsilon)\gamma_U(\epsilon)e^{i\phi_U}+\alpha^L(\epsilon)\gamma_L(\epsilon)e^{-i\phi_L})/2  &  (\epsilon-\epsilon_0)/\hbar+i(\gamma_U(\epsilon)+\gamma_L(\epsilon))/2  \\
  \end{array}
\right)^{-1},\nonumber\\
&&\label{retarded}
\end{eqnarray}
where $\bm{g}^r$ is the retarded Green's function of an isolated DQD. The advanced Green's function is $\bm{G}^a(\epsilon)=\left[\bm{G}^r(\epsilon) \right]^{\dagger}$. The linear conductance at zero temperature is given by the Green's functions and the linewidth functions at the Fermi energy $(\mu_U=\mu_L)$. In the following, we choose the Fermi energy as the origin of energy. To clarify the physical image, we consider the symmetric situation where $\gamma_U(0)=\gamma_L(0)\equiv\gamma$, $\alpha^U=\alpha^L\equiv\alpha$ and $\phi_U=\phi_L\equiv\phi/2$ except for in Sec. \ref{asymmetry}. Then, we have the following transmission probability at the Fermi energy:
\begin{eqnarray}
T(0)&=&\mbox{Tr}\left\{\bm{G}^r(0)\bm{\Gamma}^U(0)\bm{G}^a(0)\bm{\Gamma}^L(0) \right\}\nonumber\\
&=&\frac{2(\hbar\gamma)^2\left[{\epsilon_0}^2(1+\alpha^2\cos\phi)-4\alpha\epsilon_0t_c\cos\left(\frac{\phi}{2} \right)+(1+\alpha^2){t_c}^2+(\hbar\gamma)^2(1-\alpha^2)\left\{1-\alpha^2\cos^2\left(\frac{\phi}{2} \right) \right\} \right]}{\left[{\epsilon_0}^2-{t_c}^2-(\hbar\gamma)^2\left\{1-\alpha^2\cos^2\left(\frac{\phi}{2} \right) \right\} \right]^2+4(\hbar\gamma)^2\left[\epsilon_0-\alpha t_c\cos\left(\frac{\phi}{2} \right) \right]^2}.\nonumber\\
&&\label{tra}
\end{eqnarray}
This is the main result of this paper. Then, the linear conductance is given by
\begin{equation}
G(\epsilon_0,t_c,\alpha,\phi)=G_qT(0),\label{linear-conductance}
\end{equation}
where $G_q\equiv e^2/h$. This linear conductance has the symmetry of $G(-\epsilon_0,t_c,\alpha,\phi)=G(\epsilon_0,t_c,\alpha,\phi+2\pi)$. Therefore, in the following, we consider only $\epsilon_0>0$ since we can obtain the results for $\epsilon_0<0$ with adding $2\pi$ to the AB phase of the linear conductance for $\epsilon_0>0$. From the flux dependence of the conductance, we find that the phases of the AB oscillations change by $2\pi$ via the resonance except when $\alpha=0$. Similarly, we consider only $\alpha\ge 0$ since the linear conductance has the symmetry of $G(\epsilon_0,t_c,-\alpha,\phi)=G(\epsilon_0,t_c,\alpha,\phi+2\pi)$. In particular, Eq. (\ref{linear-conductance}) provides the same result obtained in Ref. \onlinecite{kang} when $\alpha=1$ and that in Ref. \onlinecite{raikh} when $\phi=0$.

\subsection{\label{linear}Energy dependence}
\begin{figure}
\includegraphics[scale=0.6]{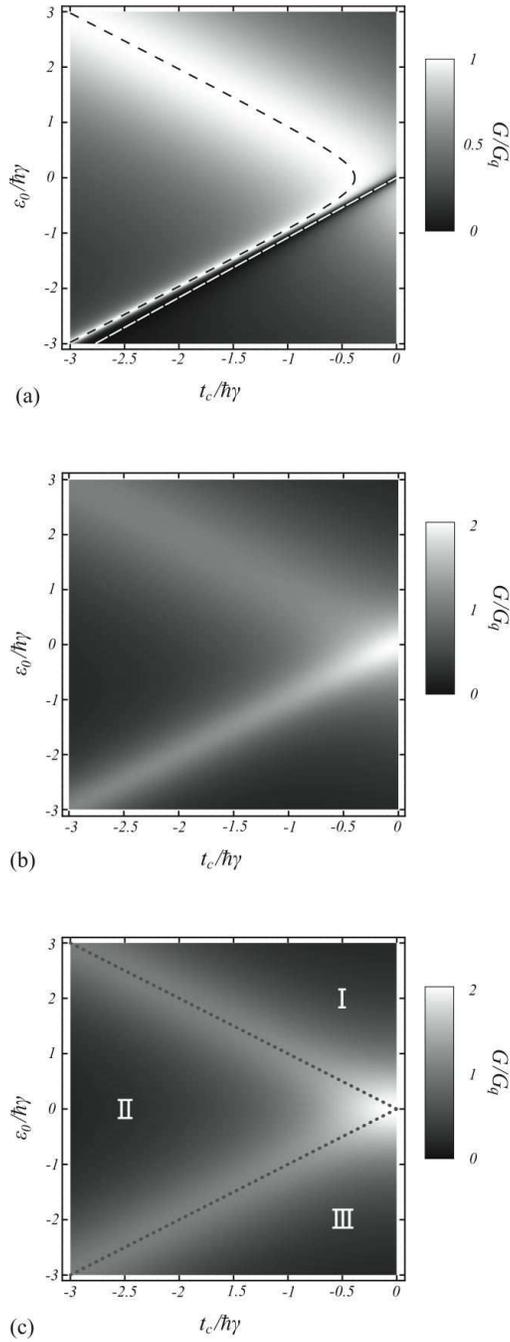}
\caption{\label{fig:linear-conductance} Gray-scale plots of the normalized linear conductance $G/G_q$ for $\Phi/\Phi_0=1/8$. (a)$\alpha=1$. (b)$\alpha=0.5$. (c)$\alpha=0$. The gray dotted lines indicate the boundaries between the weak coupling regime (regions I and III, $|\epsilon_0|>|t_c|$) and strong coupling regime (region II, $|\epsilon_0|<|t_c|$).}
\end{figure}
We define the total magnetic flux threading through the AB interferometer as $\Phi=\frac{\Phi_0}{2\pi}\phi$. In Fig. \ref{fig:linear-conductance}, we show the energy $\epsilon_0$ and the inter-dot tunnel coupling $t_c$ dependence of the linear conductance for a fixed flux $\Phi/\Phi_0=1/8$. While we use this flux value as an example, the following discussions are qualitatively valid except when $\Phi/\Phi_0$ is an integer which is discussed separately. Here we define the weak coupling regime with the condition $|\epsilon_0|>|t_c|$ and the strong coupling regime with the condition $|\epsilon_0|<|t_c|$. The former corresponds to regions I and III in Fig. \ref{fig:linear-conductance}(c), and the latter corresponds to region II in Fig. \ref{fig:linear-conductance}(c). When $\alpha=1$ (see Fig. \ref{fig:linear-conductance}(a)), there is only one mode that contributes to the conduction (see Appendix \ref{transmission}) and the conditions exist for a perfect transmission $G/G_q=1$ whose curve has the shape of hyperbolas ${\epsilon_0}^2-{t_c}^2=-(\hbar\gamma)^2\sin^2(\phi/2)$ (black broken curve in Fig. \ref{fig:linear-conductance}(a)) in the strong coupling regime and of perfect reflection $G/G_q=0$ whose curve has the shape of $\epsilon_0\cos(\phi/2)-t_c=0$ (white broken line in Fig. \ref{fig:linear-conductance}(a)) in the weak coupling regime.

When $\alpha\neq 1$, effectively two modes contribute to the conduction (see Appendix \ref{transmission}). Thus, the maximum value of $G/G_q$ may exceed $1$ (see Fig. \ref{fig:linear-conductance}(b), (c) and the changes in the scale). Moreover, at $\alpha=0$, the linear conductance becomes symmetric with respect to $\epsilon_0=0$ as shown in Fig. \ref{fig:linear-conductance}(c) since two conduction modes have transport properties through the symmetric state ($\epsilon_0+t_c$) and anti-symmetric state ($\epsilon_0-t_c$), respectively (see Eqs. (\ref{s}) and (\ref{as}) with setting $\epsilon=0$).

\begin{figure}
\includegraphics[scale=0.7]{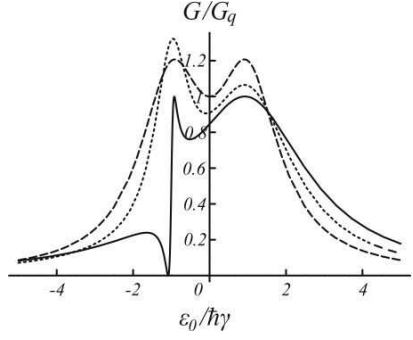}
\caption{\label{fig:fano} The $\epsilon_0$-dependences of the linear conductance when $t_c/\hbar\gamma=-1$ and $\Phi/\Phi_0=1/8$. The solid, dotted, and broken curves indicate $\alpha=1$, $0.5$, and $0$, respectively.}
\end{figure}

\begin{figure}
\includegraphics[scale=0.7]{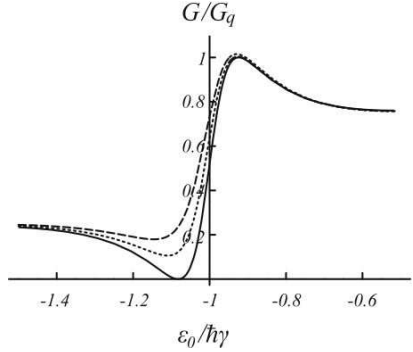}
\caption{\label{fig:fano-alpha} The suppression of Fano anti-resonance due to the indirect coupling parameter $\alpha$ when $t_c/\hbar\gamma=-1$ and $\Phi/\Phi_0=1/8$. The solid, dotted, and broken curves indicate $\alpha=1$, $0.99$, and $0.98$, respectively.}
\end{figure}

We inspect the $\epsilon_0$-dependences of the linear conductance in detail. Figure \ref{fig:fano} shows the $\epsilon_0$-dependences of the linear conductance when $t_c/\hbar\gamma=-1$ and $\Phi/\Phi_0=1/8$. When $\alpha=1$ (solid curve), the linear conductance has an asymmetric Fano lineshape near the resonance via the anti-symmetric state. The behaviors of the linear conductance at $\alpha=1$ can be explained by the Breit-Wigner resonance via the symmetric state ($\epsilon_0=-t_c$) and the Fano resonance via the anti-symmetric state ($\epsilon_0=t_c$) as discussed in Ref. \onlinecite{kang}. The Fano effect is suppressed as $\alpha$ 
decreases and the independent Breit-Wigner resonances via the symmetric and anti-symmetric states realized at $\alpha=0$ lead to two peak structures as shown in the dashed curve in Fig. \ref{fig:fano}. Although we considered $t_c/\hbar\gamma=-1$, $t_c$ merely decides the resonance positions of the symmetric and anti-symmetric states and the above discussion is valid for any finite $t_c$. In the following, we discuss the suppression of the Fano effect. When the energy of the anti-symmetric state is near the Fermi energy, the transmission probability at the Fermi energy shows the following generalized Fano form\cite{g-fano}
\begin{eqnarray}
T(0)\simeq T_b\frac{\left(\tilde{\epsilon}_a+q \right)^2+(1-\alpha)T_{\alpha}}{{\tilde{\epsilon}_a}^2+1}.\label{extended-fano}
\end{eqnarray}
Here the renormalized energy of the anti-symmetric state is defined by $\tilde{\epsilon}_a=\frac{\epsilon_a}{\hbar\Gamma_a}$. We define the linewidth functions of the symmetric and anti-symmetric states, respectively, by
\begin{equation}
\Gamma_s=\gamma\left\{1+\alpha\cos\left(\frac{\phi}{2} \right) \right\},\ \Gamma_a=\gamma\left\{1-\alpha\cos\left(\frac{\phi}{2} \right) \right\},\label{line}
\end{equation}
and the transmission probability of the background channel $T_b$ is the Breit-Wigner form via the symmetric state peaked at $\epsilon_s=\epsilon_a+2t_c\sim 2t_c$ alone
\begin{eqnarray}
T_b=\frac{1}{\left(\frac{2t_c}{\hbar\Gamma_s} \right)^2+1}.
\end{eqnarray}
The Fano parameter $q$ is positive and given by
\begin{eqnarray}
q=-\frac{2\alpha^2\sin^2\left(\frac{\phi}{2} \right)}{1-\alpha^2\cos^2\left(\frac{\phi}{2} \right)}\frac{t_c}{\hbar\Gamma_s},
\end{eqnarray}
and the additional factor $T_{\alpha}$ contributes to the transport only when $\alpha\neq 1$ and is given by
\begin{eqnarray}
T_{\alpha}=\frac{1+\alpha}{1-\alpha^2\cos^2\left(\frac{\phi}{2} \right)}\left[\left\{\frac{1-\alpha^2\cos\phi}{\alpha^2\sin^2\left(\frac{\phi}{2} \right)}q \right\}^2+1 \right].\label{26}
\end{eqnarray}
According to Eq. (\ref{extended-fano}), when $\alpha=1$, the transmission probability at the Fermi energy provides the Fano lineshape and reproduces the result of Ref. \onlinecite{kang}. Moving away from $\alpha=1$, however, the transmission probability deviates from the Fano lineshape because of the additional term in Eq. (\ref{extended-fano}). Therefore, the Fano effect is suppressed and the Fano anti-resonance disappears as shown in detail in Fig. \ref{fig:fano-alpha}. For smaller $|\phi|$, the contribution of $T_{\alpha}$ is larger as seen in Eq. (\ref{26}). Although here we discuss the situation of $\Phi/\Phi_0\simeq 2n$, where $n$ is an integer, the role of the symmetric and anti-symmetric states are interchanged when $\Phi/\Phi_0\simeq (2n+1)$ and the Fano parameter is negative and given by
\begin{equation}
q=\frac{2\alpha^2\sin^2\left(\frac{\phi}{2} \right)}{1-\alpha^2\cos^2\left(\frac{\phi}{2} \right)}\frac{t_c}{\hbar\Gamma_a}.
\end{equation}

\begin{figure}
\includegraphics[scale=0.4]{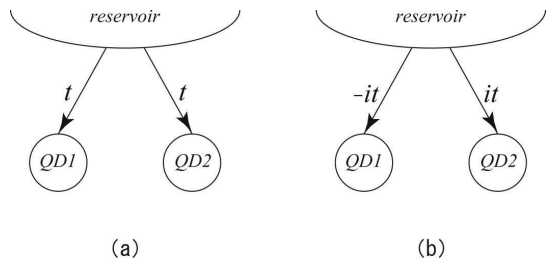}
\caption{\label{fig:couple} The tunnel coupling between reservoirs and QDs. The tunneling parameter $t$ represents $t_{\nu k}^{(i)}$ in Eq. (\ref{bardeen-form}). (a) The tunnel coupling between reservoirs and QDs at $\Phi/\Phi_0=2n$. (b) The tunnel coupling between reservoirs and QDs at $\Phi/\Phi_0=2n+1$.}
\end{figure}

\begin{figure}
\includegraphics[scale=0.7]{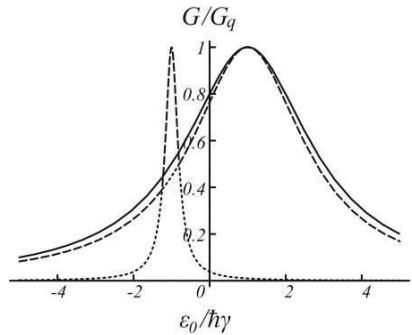}
\caption{\label{fig:alpha-effect} The linear conductance as a function of single electron energy level $\epsilon_0$. The solid curve indicates the linear conductance for $\alpha=1$, $t_c/\hbar\gamma=-1$, and $\Phi/\Phi_0=2n$. The dotted, and broken curves indicate the linear conductances of two modes $T_1$ and $T_2$ defined by Eqs. (\ref{t1}) and (\ref{t2}), respectively, for $\alpha=0.8$, $t_c/\hbar\gamma=-1$, and $\Phi/\Phi_0=2n$.}
\end{figure}

Then we discuss the special situation that arises when $\Phi/\Phi_0$ is an integer. First, we consider $\alpha=1$ where only a single conduction mode contributes to the transport. When $\Phi/\Phi_0=2n$ electrons can only couple with the symmetric state since electrons tunnel from reservoirs into QDs with the same tunnel coupling under this condition as shown in Fig. \ref{fig:couple}(a). Similarly, when $\Phi/\Phi_0=2n+1$, electrons can only couple with the anti-symmetric state (see Fig. \ref{fig:couple}(b)). We consider the effects of $\alpha$ on the energy dependence of the linear conductance when $\Phi/\Phi_0=2n$. When $\alpha=1$, the linear conductance exhibits the lineshape of a Breit-Wigner type transmission via the symmetric state as shown in Fig. \ref{fig:alpha-effect}. However, when $\alpha\neq 1$, the linear conductance has a side peak structure via the anti-symmetric state in addition to the Breit-Wigner like transmission via the symmetric state. This is because transport becomes possible by both the symmetric and the anti-symmetric states. The linewidth functions in the symmetric and anti-symmetric base are
\begin{eqnarray}
\bm{\Gamma}^U=\bm{\Gamma}^L=\left(
  \begin{array}{cc}
    (1+\alpha)\gamma   & 0   \\
    0   & (1-\alpha)\gamma   \\
  \end{array}
\right).
\end{eqnarray}
This is equivalent to Eq. (\ref{line}) with $\phi=4n\pi$. Thus, electrons can couple with both the symmetric and anti-symmetric states when $\alpha\neq 1$. The ratio of the full-width-at-half-maximum of the two conductance peaks shown in Fig. \ref{fig:alpha-effect} is the ratio of $1+\alpha$ and $1-\alpha$. Similarly, when $\Phi/\Phi_0=2n+1$, the roles of the symmetric and anti-symmetric states are reversed.

\subsection{\label{ab}AB oscillations}
In this section, we discuss the magnetic flux dependence of the linear conductance. In particular, the linear conductance is independent of the flux at $\alpha=0$ and the visibility of AB oscillation (i.e., the ratio of the AB oscillation amplitude and the current maximum) becomes zero in all situations.

\subsubsection{Zero coupling}
\begin{figure}
\includegraphics[scale=0.7]{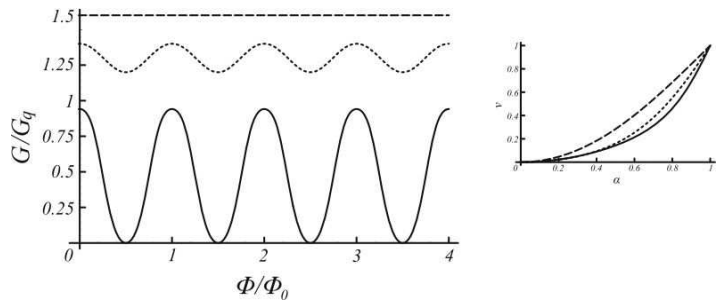}
\caption{\label{fig:no-tunnel-coupling} AB oscillations in the absence of inter-dot tunnel coupling. The solid, dotted, and broken curves indicate $\alpha=1$, $0.5$, and $0$, respectively, when $\epsilon_0/\hbar\gamma=0.5$. Inset: $\alpha$ dependence of the visibility of zero coupling. The solid, dotted, and broken curves indicate $\epsilon_0/\hbar\gamma=0.5$, $1$, and $2$, respectively.}
\end{figure}

First, we consider the weak coupling limit, namely the absence of direct tunneling, $t_c=0$. AB oscillations are shown in Fig. \ref{fig:no-tunnel-coupling}. When $\alpha=1$, the linear conductance shows a usual AB oscillation with a visibility of $1$ and a period of $2\pi$ because of the effect of the interference between the probability amplitude associated with one path through QD 1 and the probability amplitude associated with another path through QD 2.

For general $\alpha$, the visibility $v$ is given by
\begin{eqnarray}
v=1-\frac{(1-\alpha^2)\left[\left\{{\epsilon_0}^2-(\hbar\gamma)^2(1-\alpha^2) \right\}^2+4(\hbar\gamma)^2{\epsilon_0}^2 \right]}{\left[{\epsilon_0}^2+(\hbar\gamma)^2 \right]\left[{\epsilon_0}^2(1+\alpha^2)+(\hbar\gamma)^2(1-\alpha^2)^2 \right]}
\end{eqnarray}
The indirect coupling parameter dependences of the visibility is shown in inset to Fig. \ref{fig:no-tunnel-coupling}. The visibility decreases as $\alpha$ decreases. In particular, when $|\epsilon_0/\hbar\gamma|\gg 1$, $v\simeq 2\alpha^2/(1+\alpha^2)$. While, when $|\epsilon_0/\hbar\gamma|\ll 1$, $v\simeq \alpha^2$. For any $\epsilon_0$, the visibility $v$ is monotonically increasing function of $\alpha$.

\subsubsection{Weak coupling regime}
\begin{figure}
\includegraphics[scale=0.7]{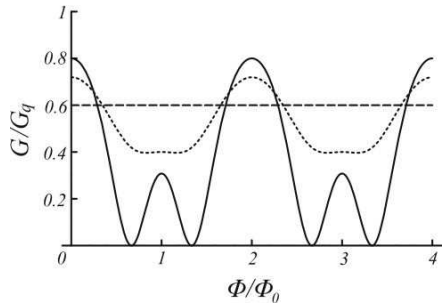}
\caption{\label{fig:weak-coupling} AB oscillations in the weak coupling regime, $\epsilon_0/\hbar\gamma=2$ and $t_c/\hbar\gamma=-1$. The solid, dotted, and broken curves indicate $\alpha=1$, $0.5$, and $0$, respectively.}
\end{figure}

Here we consider the weak coupling regime, $|\epsilon_0|>|t_c|$. AB oscillations in this regime are shown in Fig. \ref{fig:weak-coupling} for various parameters $\alpha$. There are completely destructive interferences for $\alpha=1$.
The visibility decreases as $\alpha$ decreases as found in the weak coupling limit. 
The AB oscillation period becomes $4\pi$ since the linear conductance has an alternating peak structure via the symmetric and anti-symmetric states because of the finite inter-dot tunnel coupling. In this situation, the peak heights of the conductance via the symmetric state are higher than those via the anti-symmetric state since the energy of the latter is further from Fermi energy than that of the former.

\subsubsection{\label{scr}Strong coupling regime}
\begin{figure}
\includegraphics[scale=0.7]{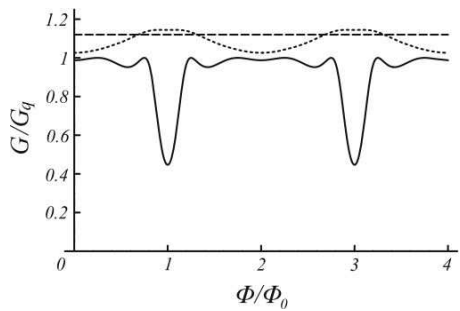}
\caption{\label{fig:strong-coupling} AB oscillations in the strong coupling regime, $\epsilon_0/\hbar\gamma=1$ and $t_c/\hbar\gamma=-\sqrt{3/2}$. The solid, dotted, and broken curves indicate $\alpha=1$, $0.5$, and $0$, respectively.}
\end{figure}

Next, we consider the strong coupling regime, $|\epsilon_0|<|t_c|$. When $\alpha=1$, there is one conduction mode and the perfect transmission $G/G_q=1$ is realized. For the parameters shown in Fig. \ref{fig:strong-coupling}, the resonant condition (${\epsilon_0}^2-{t_c}^2=-(\hbar\gamma)^2\sin^2(\phi/2)$) is satisfied at $\Phi/\Phi_0=(2n+1)/4$, where $n$ is an integer. Unlike to the weak coupling regime, there are no completely destructive interferences in the AB oscillations. The visibility decreases as 
$\alpha$ decreases, which is similar to the results for the weak coupling regime.

It is interesting to note that the conductance is maximum at integer $\Phi/\Phi_0$ for weak coupling, while the conductance is minimum at that flux for strong coupling. This is more evident by comparing the data at $\alpha=0.5$ in Figs. \ref{fig:weak-coupling} and \ref{fig:strong-coupling}. Such a change of the AB oscillation phase by $t_c$ bears intriguing analogy to the AB oscillation phase change by the energy difference between the two QDs pointed in Ref. \onlinecite{kubala}.

\subsubsection{Double resonant levels}
\begin{figure}
\includegraphics[scale=0.7]{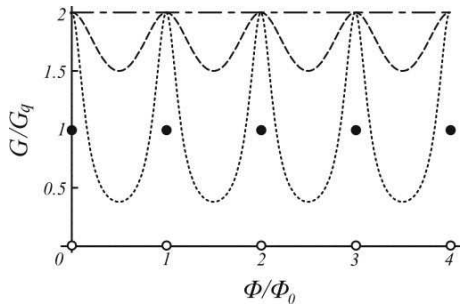}
\caption{\label{fig:double-resonance} AB oscillations in the double resonant levels, $\epsilon_0/\hbar\gamma=t_c/\hbar\gamma=0$. The solid, dotted, and broken, and dash-dotted curves indicate $\alpha=1$, $0.9$, $0.5$, and $0$, respectively.}
\end{figure}

Here we consider a special case when $\epsilon_0=t_c=0$, namely two energy levels of the DQD are aligned with the Fermi level (resonant levels). The linear conductance shows a singular flux dependence at $\alpha=1$ as discussed in Refs. \onlinecite{kubala} and \onlinecite{raikh} (see Fig. \ref{fig:double-resonance}), namely the linear conductance becomes zero except for $\Phi/\Phi_0=n$ where $G=G_q$ is realized. As in the previous sections, the visibility decreases as $\alpha$ decreases and finally the perfect transmission ($G/G_q=2$) is realized at $\alpha=0$ since two conduction modes satisfy the resonant condition ($\epsilon_0=0$).

\section{\label{asymmetry}Tunneling current in an asymmetric AB interferometer}
In this section, we discuss the tunneling current in an asymmetric AB interferometer, namely $\gamma_U\neq \gamma_L$ or $\phi_U\neq\phi_L$. However, it is well-known that the asymmetry of the linewidth functions suppresses the Breit-Wigner resonance because of the asymmetric factor $\Gamma^U\Gamma^L/(\Gamma^U+\Gamma^L)$. Therefore, in this section, we only study the asymmetry of the flux. As in the previous sections, there are two conduction modes in the asymmetric systems except when $\alpha= 1$ where only a single conduction mode contributes to the transport. The behavior of the Fano effect due to the asymmetry of the flux has been already investigated\cite{fano-as}. In their analysis of $\alpha=1$, the conductance can be expressed as the Fano form with flux-imbalance dependent energy scaling and Fano parameter $q$. We studied the effect of $\alpha$ less than $1$ and found that the Fano resonance is very sensitive to the parameter $\alpha$ similar to the results in Sec. \ref{linear} (not shown). Here we only show the AB oscillation results with several $\alpha$ parameters emphasizing the difference for regimes of interest.

In the situation of zero coupling ($t_c=0$), only the total flux influences the relative phases of the electrons tunneling through the two dots since there is no direct tunneling between dots. Thus, the visibility of the AB oscillations is independent of the asymmetry of the flux in the weak coupling limit.

\subsection{Weak coupling regime}
\begin{figure}
\includegraphics[scale=0.7]{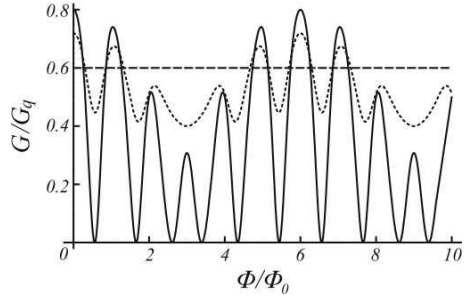}
\caption{\label{fig:generalization-weak-coupling-flux} $\gamma_U=\gamma_L=\gamma$, $\Phi_U/\Phi_0=\Phi/6\Phi_0$, $\Phi_L/\Phi_0=5\Phi/6\Phi_0$, $\epsilon_0/\hbar\gamma=2$, and $t_c/\hbar\gamma=-1$. The solid, dotted, and broken curves indicate $\alpha=1$, $0.5$, and $0$, respectively.}
\end{figure}

Here we consider the weak coupling regime, $|\epsilon_0|>|t_c|$. The AB oscillations for $\Phi_U=1/48$ and $\Phi_L=5/48$ are shown in Fig. \ref{fig:generalization-weak-coupling-flux}.
When we compare Figs. \ref{fig:weak-coupling} and \ref{fig:generalization-weak-coupling-flux}, the period of the AB oscillations changes due to the asymmetry of the flux. The periods of the AB oscillations are determined by the flux threading through a smaller subcircuit. In Fig. \ref{fig:generalization-weak-coupling-flux}, the flux threading through a smaller subcircuit is $\Phi/6\Phi_0$, and the period is $12\pi$. When the ratio of the flux is an irrational number, the AB oscillations no longer have a well-defined period\cite{aharony}. For $|\alpha|<1$, the visibility of the AB oscillation decreases as the $|\alpha|$ decreases as well as the symmetric situation.

\subsection{Strong coupling regime}
\begin{figure}
\includegraphics[scale=0.7]{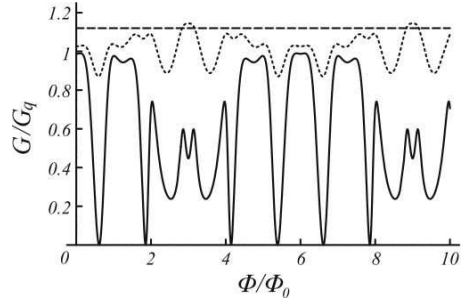}
\caption{\label{fig:generalization-strong-coupling-flux} $\gamma_U=\gamma_L=\gamma$, $\Phi_U/\Phi_0=\Phi/6\Phi_0$, $\Phi_L/\Phi_0=5\Phi/6\Phi_0$, $\epsilon_0/\hbar\gamma=1$, and $t_c/\hbar\gamma=-\sqrt{3/2}$. The solid, dotted, and broken curves indicate $\alpha=1$, $0.5$, and $0$, respectively.}
\end{figure}

Here we consider the strong coupling regime, $|\epsilon_0|<|t_c|$. The AB oscillations for $\Phi_U=1/48$ and $\Phi_L=5/48$ are shown in Fig. \ref{fig:generalization-strong-coupling-flux}.
When we compare Figs. \ref{fig:strong-coupling} and \ref{fig:generalization-strong-coupling-flux}, the period of the AB oscillations changes due to the asymmetry of the flux. The periods of the AB oscillations are determined by the flux threading through a smaller subcircuit as found in the weak coupling regime. Unlike the situation of the symmetric flux as discussed in Sec. \ref{scr} where completely destructive interference was absent, we could observe sharp zero conductance dips when $\alpha=1$. The visibility of the AB oscillation decreases as the $\alpha$ decreases, which is similar to the symmetric situation.

\section{\label{conclusion}Conclusion}
In this paper, we investigated the electron transport through a DQD system with the indirect coupling parameter $\alpha$. We found that there are two conduction modes except when $|\alpha|=1$. The Fano anti-resonance is suppressed for $|\alpha|<1$. The visibility of AB oscillations decreases as $|\alpha|$ decreases and the visibility reaches zero when $\alpha=0$ since the tunneling conductance is independent of the flux. Moreover, we have investigated the effects of the asymmetry of the flux and the tunnel coupling between reservoirs and dots. The asymmetry of the flux leads to a change in the period of the AB oscillations.

\begin{acknowledgments}
We thank B. L. Altshuler, G. E. W. Bauer, U. Zuelicke, A. Aharony, O. Entin-Wohlman, and M. Pioro-Ladri\`{e}re for valuable discussions and useful comments. One of the authors (Y. T.) is partly supported by SORST-JST.
\end{acknowledgments}

\appendix

\section{\label{dicp}Derivation of indirect coupling parameter $\alpha$}
Here we show the detailed derivation of the indirect coupling parameter $\alpha$ discussed in Section \ref{icp}. For the model given in Section \ref{icp}, the wave functions of reservoir modes and their eigenvalues are
\begin{eqnarray}
\phi_{Uk}(r)&=&\frac{\sqrt{2}}{L^{d/2}}e^{i\vec{k_{\parallel}}\cdot\vec{\rho}}\sin\left(\frac{\theta}{2}\right)e^{-\kappa y},\\
\epsilon_{Uk}&=&\frac{\hbar{k_{\parallel}}^2}{2\mu}+U-\frac{\hbar^2\kappa^2}{2\mu},
\end{eqnarray}
where the decay constant $\kappa$ is
\begin{eqnarray}
\kappa=\frac{1}{\hbar}\sqrt{2\mu\left(U-\frac{\hbar^2{k_y}^2}{2\mu}\right)},
\end{eqnarray}
and the scattering phase shift $\theta$ satisfies $\sin\theta=\frac{2\kappa k_y}{\kappa^2+{k_y}^2}$ derived by matching wave functions $\zeta (y)=e^{ik_yy}+re^{-ik_yy}$ for $y<0$ and $\zeta(y)=Ce^{-\kappa y}$ for $y>0$. We found $r=-e^{i\theta}$ and $C=-2ie^{i\frac{\theta}{2}}\sin\frac{\theta}{2}$. $L$ is the system size and $d$ is the dimensionality of the reservoir, three-dimension (3D) ($\vec{k_{\parallel}}=(k_x,k_z)$ and $\vec{\rho}=(x,z)$) or two-dimension (2D) ($\vec{k_{\parallel}}=(k_x,0)$ and $\vec{\rho}=(x,0)$). For the dot, we consider only the symmetric mode, namely zero angular momentum state,
\begin{eqnarray}
\phi_d(r>R_0)=\left\{
  \begin{array}{cc}
    D_3\frac{1}{\sigma r}e^{-\sigma r}   & (3D),   \\
    D_2K_0(\sigma r)   &  (2D),  \\
  \end{array}
\right.
\end{eqnarray}
where $\sigma^2+{k_{\parallel}}^2=\kappa^2$ and $D_d$ is a constant related to the detail of the potential inside the dot and $K_0$ is the $0$th order modified Bessel function. The origin should be shifted to the center of each dot $\vec{r}_i$. To calculate the tunneling matrix element, it is convenient to obtain the Fourier transform of them, for 3D:
\begin{eqnarray}
\phi_d^{(i)}(r)=\frac{2\pi D_3}{\sigma}\int\frac{d^2q}{(2\pi)^2}\frac{e^{-\sqrt{\sigma^2+q^2}(y_i-y)}}{\sqrt{\sigma^2+q^2}}e^{-i(q_x(x-x_i)+q_zz)},
\end{eqnarray}
and for 2D:
\begin{eqnarray}
\phi_d^{(i)}(r)=\pi D_2\int\frac{dq_x}{2\pi}\frac{e^{-\sqrt{\sigma^2+{q_x}^2}(y_i-y)}}{\sqrt{\sigma^2+{k_x}^2}}e^{-ik_x(x-x_i)}.
\end{eqnarray}
Using Eq. (\ref{bardeen-form}) for 3D system, the tunneling matrix elements are
\begin{eqnarray}
t_{Uk}^{(i)}=\frac{\hbar^2}{2\mu}\frac{\sqrt{2}}{L^{3/2}}\sin\left(\frac{\theta}{2} \right)\frac{4\pi D_3}{\sigma}e^{-\kappa y_i+ik_xx_i},
\end{eqnarray}
and for 2D system are
\begin{eqnarray}
t_{Uk}^{(i)}=\frac{\hbar^2}{2\mu}\frac{\sqrt{2}}{L}\sin\left(\frac{\theta}{2} \right)2\pi D_2e^{-\kappa y_i+ik_xx_i}.
\end{eqnarray}
Therefore, using the dimensionality-dependent constant $A_d$, we have in general
\begin{eqnarray}
t_{Uk}^{(i)}=A_d\sin\left(\frac{\theta}{2} \right)e^{-\kappa y_i+ik_xx_i}.
\end{eqnarray}

Using Eq. (\ref{g}) for 3D system, the general formula of linewidth functions at the Fermi energy $\mu_U\equiv\frac{\hbar^2{k_F}^2}{2\mu}$ is
\begin{eqnarray}
\Gamma_{ij}^U(\mu_U)=\frac{(2\pi)^2{A_3}^2}{U}\left(\frac{L}{2\pi} \right)^3\int_{0}^{k_F}dk\cdot k\sqrt{{k_F}^2-k^2}e^{-\sqrt{\sigma^2+k^2}(y_i+y_j)}J_0(ks),
\end{eqnarray}
where $s=|x_i-x_j|$, which have a limiting form for $U\gg \mu_U$,
\begin{eqnarray}
\Gamma_{ij}^U(\mu_U)\sim\frac{(2\pi)^2{A_3}^2}{U}\left(\frac{L}{2\pi} \right)^3e^{-(y_i+y_j)\sigma}\frac{{k_F}^2}{s}j_1(k_Fs),
\end{eqnarray}
and for $U\gtrsim \mu_U$,
\begin{eqnarray}
\Gamma_{ij}^U(\mu_U)\sim\frac{(2\pi)^2{A_3}^2}{U}\left(\frac{L}{2\pi} \right)^3\frac{k_F(y_i+y_j)}{(s^2+(y_i+y_j)^2)^{3/2}}.
\end{eqnarray}
Similarly, the formula for 2D system is
\begin{eqnarray}
\Gamma_{ij}^U(\mu_U)=\frac{2\pi {A_2}^2}{U}\left(\frac{L}{2\pi} \right)^2\int_{-k_F}^{k_F}dk_x\sqrt{{k_F}^2-{k_x}^2}e^{-\sqrt{\sigma^2+{k_x}^2}(y_i+y_j)+ik_xs},
\end{eqnarray}
and a limiting form for $U\gg \mu_U$,
\begin{eqnarray}
\Gamma_{ij}^U(\mu_U)\sim\frac{2\pi {A_2}^2}{U}\left(\frac{L}{2\pi} \right)^2\frac{\pi k_F}{s}e^{-\sigma(y_i+y_j)}J_1(k_Fs),
\end{eqnarray}
and for $U\gtrsim \mu_U$,
\begin{eqnarray}
\Gamma_{ij}^U(\mu_U)\sim\frac{2\pi {A_2}^2}{U}\left(\frac{L}{2\pi} \right)^2\frac{2k_F(y_i+y_j)}{s^2+(y_i+y_j)^2}.
\end{eqnarray}

We define the indirect coupling parameter as follows:
\begin{eqnarray}
\alpha^U(\epsilon)\equiv\frac{\Gamma_{12}^U(\epsilon)}{\sqrt{\Gamma_{11}^U(\epsilon)\Gamma_{22}^U(\epsilon)}}.
\end{eqnarray}
Therefore, the parameter at the Fermi energy for 3D system with $U\gg \mu_U$,
\begin{eqnarray}
\alpha^U(\mu_U)\sim\frac{3}{k_Fs}j_1(k_Fs),
\end{eqnarray}
where $j_1$ is the first order spherical Bessel function and for $U\gtrsim \mu_U$
\begin{eqnarray}
\alpha^U(\mu_U)\sim\frac{(2y_d)^3}{(s^2+(2y_d)^2)^{3/2}},
\end{eqnarray}
where we consider symmetric configuration: $y_1=y_2=y_d$. For 2D system with $U\gg \mu_U$,
\begin{eqnarray}
\alpha^U(\mu_U)\sim\frac{2}{k_Fs}J_1(k_Fs),
\end{eqnarray}
where $J_1$ is the first order Bessel function and for $U\gtrsim \mu_U$
\begin{eqnarray}
\alpha^U\sim\frac{(2y_d)^2}{s^2+(2y_d)^2},
\end{eqnarray}
where we consider symmetric configuration. In Ref. \onlinecite{raikh}, although similar derivation of the parameter $\alpha$ of 2D system was discussed, their result ($\alpha\propto J_0(k_Fs)$) is different from ours. Our result is more general since the authors of Ref. \onlinecite{raikh} ignored $k_y$ dependence of the  scattering phase shift.

\section{\label{transmission}Transmission probability}
We diagonalize the transmission matrix at the Fermi energy $\bm{T}(0)=\bm{G}^r(0)\bm{\Gamma}^U(0)\bm{G}^a(0)\bm{\Gamma}^L(0)$ found in Eq. (\ref{tra}). We then obtain two eigenvalues of the transmission matrix
\begin{eqnarray}
T_1(\epsilon)&=&\frac{(\hbar\gamma)^2\left[T_A(\epsilon)- T_B(\epsilon)\right]}{\left[(\epsilon-\epsilon_0)^2-{t_c}^2-(\hbar\gamma)^2\left\{1-\alpha^2\cos^2\left(\frac{\phi}{2} \right) \right\} \right]^2+4(\hbar\gamma)^2\left[(\epsilon-\epsilon_0)+\alpha t_c\cos\left(\frac{\phi}{2} \right) \right]^2},\label{t1}\\
T_2(\epsilon)&=&\frac{(\hbar\gamma)^2\left[T_A(\epsilon)+ T_B(\epsilon)\right]}{\left[(\epsilon-\epsilon_0)^2-{t_c}^2-(\hbar\gamma)^2\left\{1-\alpha^2\cos^2\left(\frac{\phi}{2} \right) \right\} \right]^2+4(\hbar\gamma)^2\left[(\epsilon-\epsilon_0)+\alpha t_c\cos\left(\frac{\phi}{2} \right) \right]^2},\label{t2}
\end{eqnarray}
with the two positive functions
\begin{eqnarray}
T_A(\epsilon)&=&(\epsilon-\epsilon_0)^2(1+\alpha^2\cos\phi)+4\alpha(\epsilon-\epsilon_0)t_c\cos\left(\frac{\phi}{2} \right)+(1+\alpha^2){t_c}^2+(\hbar\gamma)^2(1-\alpha^2)\left\{1-\alpha^2\cos^2\left(\frac{\phi}{2} \right) \right\},\nonumber\\
&&\\
T_B(\epsilon)&=&2\left|\alpha(\epsilon-\epsilon_0)\cos\left(\frac{\phi}{2} \right)+t_c \right|\nonumber\\
&&\times\sqrt{\alpha^2(\hbar\gamma)^2(1-\alpha^2)\sin^2\left(\frac{\phi}{2} \right)+(\epsilon-\epsilon_0)^2\left\{1-\alpha^2\sin^2\left(\frac{\phi}{2} \right)\right\}+2\alpha(\epsilon-\epsilon_0)t_c\cos\left(\frac{\phi}{2} \right)+\alpha^2{t_c}^2}.
\end{eqnarray}

When $|\alpha|=1$,
\begin{eqnarray}
T_A(\epsilon)=T_B(\epsilon)=2\left\{(\epsilon-\epsilon_0)\cos\left(\frac{\phi}{2} \right)+t_c \right\}^2,
\end{eqnarray}
and the transmission probabilities are
\begin{eqnarray}
&&T_1(\epsilon)=0,\\
&&T_2(\epsilon)=\frac{4(\hbar\gamma)^2\left[(\epsilon-\epsilon_0)\cos\left(\frac{\phi}{2} \right)+t_c \right]^2}{\left[(\epsilon-\epsilon_0)^2-{t_c}^2-(\hbar\gamma)^2\sin^2\left(\frac{\phi}{2} \right) \right]^2+4(\hbar\gamma)^2\left[(\epsilon-\epsilon_0)+t_c\cos\left(\frac{\phi}{2} \right) \right]^2}.
\end{eqnarray}
Therefore, we have single conduction mode at $|\alpha|=1$. When $|\alpha|<1$, $T_A\neq T_B$ and the number of the conduction modes is two. In fact, we have the following relation:
\begin{eqnarray}
{T_A}^2-{T_B}^2=(1-\alpha^2)^2\left[1+\tilde{\alpha}(2-\tilde{\alpha})+(\epsilon-\epsilon_0-t_c)^2 \right]\left[1-\tilde{\alpha}(2-\tilde{\alpha})+(\epsilon-\epsilon_0+t_c)^2 \right]>0,
\end{eqnarray}
where $\tilde{\alpha}=\alpha\cos\left(\frac{\phi}{2} \right)$.

In particular, when $\alpha=0$, the transmission probabilities are
\begin{eqnarray}
&&T_1(\epsilon)=\frac{(\hbar\gamma)^2}{(\epsilon-\epsilon_0-t_c)^2+(\hbar\gamma)^2},\label{s}\\
&&T_2(\epsilon)=\frac{(\hbar\gamma)^2}{(\epsilon-\epsilon_0+t_c)^2+(\hbar\gamma)^2}.\label{as}
\end{eqnarray}
$T_1$ and $T_2$ represent Breit-Wigner resonances through the symmetric and anti-symmetric states, respectively. These are independent of the flux.

\newpage 


\begin{thebibliography}{20}
\bibitem{offset} In this paper, we focus on the effects of $\alpha$ which characterizes the coherent coupling via the reservoirs between two quantum dots. Then, since it is physically interesting to compare that with the effects of the direct coherent coupling $t_c$ between two quantum dots, we had given priority to the effects of $t_c$ rather than that of the energy difference. Therefore, we assume that the energies of the two QDs are equal.
\bibitem{g-fano} The generalized Fano form had been initially introduced in Ref. \onlinecite{kubala} in a different context.
\end{thebibliography}
\end{document}